\documentclass[journal]{IEEEtran}

\IEEEoverridecommandlockouts           
\usepackage{epsfig} 
\usepackage{amsmath,amssymb,amsfonts}
\usepackage{color}
\usepackage{lipsum}
\usepackage{balance}
\usepackage{booktabs,tabularx,ifsym}
\usepackage{eurosym}
\usepackage{multirow}
\usepackage{mathtools}
\usepackage{pdfrender}
\usepackage{trfsigns}

\usepackage{centernot}
\usepackage{color}
\usepackage[acronym]{glossaries}
\usepackage[ruled]{algorithm2e}
\usepackage{graphicx}
\usepackage{wrapfig}
\usepackage{tabu}
\usepackage{subfig}

\graphicspath{{images/}}

\newcommand{\R}{\mathbb{R}}

\newcommand{\B}{\mathbb{B}}

\newcommand{\N}{\mathbb{N}}

\newcommand{\mc}[1]{\mathcal{#1}}

\DeclareMathOperator*{\argmin}{argmin}

\newcommand{\until}[1]{\{1,\dots,#1\}}

\newcommand{\col}{\mathrm{col}}
\newcommand{\tup}[1]{\textup{#1}}
\newcommand{\bs}[1]{\boldsymbol{#1}}
\newcommand{\QEDopen}{\hfill $\square$}

\newtheorem{theorem}{Theorem}

\newtheorem{definition}{Definition}
\newtheorem{proposition}{Proposition}

\newtheorem{remark}{Remark}

\makeglossaries
\newacronym{MIP}{MIP}{Mixed-Integer Programming}
\newacronym{SoC}{SoC}{State of Charge}
\newacronym{PEV}{PEV}{Plug-in Electric Vehicle}
\newacronym{EV}{EV}{Electric Vehicle}
\newacronym{MLD}{MLD}{Mixed-Logical-Dynamical}
\newacronym{GS}{GS}{Gauss-Southwell}
\newacronym{GNEP}{GNEP}{Generalized Nash Equilibrium Problem}
\newacronym{MI-GPG}{MI-GPG}{Mixed-Integer Generalized Potential Game}
\newacronym{MINE}{$\varepsilon$-MINE}{$\varepsilon$-Mixed-Integer Nash Equilibrium}
\newacronym{CTM}{CTM}{Cell Transmission Model}
\newacronym{CS}{CS}{Charging Station}
\newacronym{FV}{FV}{fuel vehicle}
\newacronym{r2s}{$\mathrm{r2s}$}{road-to-station}
\newacronym{s2r}{$\mathrm{s2r}$}{station-to-road}
\newacronym{HO}{HO}{Highway Operator}
\newacronym{TDM}{TDM}{Traffic Demand Management}
\newacronym{ATDM}{ATDM}{Active Traffic Demand Management}

\begin{document}
%
\title{Highway Traffic Control via\\ Smart e-Mobility -- Part I: Theory}
%
%
%

\author{Carlo Cenedese$^{\textrm{(a)}}$, Michele Cucuzzella $^{\textrm{(b),\,(c)}}$, Jacquelien M. A. Scherpen$^{\textrm{(c)}}$,\\Sergio Grammatico$^{\textrm{(d)}}$ and  Ming Cao$^{\textrm{(c)}}$  
\thanks{$^{\textrm{(a)}}$ Department of Information Technology and Electrical Engineering, ETH Z\"urich, Zurich, Switzerland ({\texttt{ccenedese@ethz.ch}}).
$^{\textrm{(b)}}$ 
Department of Electrical, Computer and Biomedical Engineering, University of Pavia, Pavia, Italy ({\texttt{michele.cucuzzella@unipv.it}}).
$^{\textrm{(c)}}$ Jan C. Willems Center for Systems and Control, ENTEG, Faculty of Science and Engineering, University of Groningen, The Netherlands	({\texttt{\{j.m.a.scherpen, m.cao\}@rug.nl}}). 
$^{\textrm{(d)}}$ Delft Center for Systems and Control, TU Delft, The Netherlands
	({\texttt{s.grammatico@tudelft.nl}}). 	
	The work of Cenedese and Cao was supported by The Netherlands Organization for Scientific Research (NWO-vidi-14134), the one of Cucuzzella and Scherpen by the EU Project \lq MatchIT' (82203), and the one of Grammatico by NWO under project OMEGA (613.001.702) and P2P-TALES (647.003.003) and by the ERC under research project COSMOS (802348).}}

\markboth{IEEE - Transaction on Intelligent Transportation Systems, \today}%
{Shell \MakeLowercase{\textit{et al.}}: Bare Demo of IEEEtran.cls for IEEE Journals}

\maketitle

\begin{abstract}
In this paper, we study how to alleviate highway traffic congestion by encouraging plug-in hybrid and electric  vehicles to stop at a charging station around peak congestion times. Specifically, we design a pricing policy to make the charging price dynamic and dependent on the traffic congestion, predicted via the cell transmission model, and the availability of charging spots. Furthermore, we develop a novel framework to model how this policy affects the drivers' decisions by formulating a mixed-integer potential game. Technically, we introduce the concept of ``road-to-station'' (r2s) and ``station-to-road'' (s2r) flows, and  show that the selfish actions of the drivers converge to charging schedules that are individually optimal in the sense of Nash. In the second part of this work, submitted as a separate paper (Part II: Case Study), we validate the proposed strategy on a simulated  highway stretch between The Hague and Rotterdam, in The Netherlands.
\end{abstract}


%
\IEEEpeerreviewmaketitle

\section{Introduction}

\subsection{Motivation}
\IEEEPARstart{I}{n} the recent years, urban mobility in  highly populated cities is becoming a central issue  in many countries. Some alarming statistics show a pressing need for change, as the cost of congestion to the EU society is no less than \euro$ 267$ billion per  year~\cite{eu:2019:traffic}. In fact, an inefficient transportation system deteriorates not only the citizens' well-being, but also the environment, since traffic jams heavily increase the emission of $\text{CO}_2$ \cite{barth:2009:traffic}.
The classical solution to the \gls{TDM} problem is to increase the roads' capacity or to build alternative routes. Although  this approach produces tangible benefits~\cite{ganine:2017:resilience},
  policymakers and researchers are exploring alternatives that may be sensibly faster  and cheaper to implement, and provide  dynamic solutions that adapt to the traffic evolution.
\subsection{``Hard'' and ``soft'' policies}
In the past years, there has been a growing interest from the research community in the problem of \gls{ATDM}, i.e., a dynamic  or even real-time solution to the traffic control problem. The literature on the topic can be divided into the design of ``hard'' and ``soft''  policies to address the problem~\cite{goodwin:2008:traffic_soft_measures}.  The hard type of measures tries to enforce changes in the  drivers' behavior by imposing some constraints or penalizing undesired actions. For example, several works studied the use of dynamic traffic signaling or traffic lights to influence the current traffic flow~\cite{como:2016:dynamic_signaling,guo:2019:signal_control_survey}. In \cite{piacentini:ferrara:2018:moving_bottleneck}, the authors impose an artificial bottleneck to decrease the flow in strategic part of the road and achieve an alleviation of the congestion. Another approach is to increase the transit price of the most congested roads in order to boost the use of alternative routes~\cite{downs:2005:still_stuck_in_traffic}.

On the other hand, the so-called soft measures are designed to incentivize virtuous driver behaviors, and have their roots in behavioural economics and psychology. The word \emph{soft} refers to the possibility of the  drivers to ignore the incentives and stick to  their regular conduct~\cite{goodwin:2008:traffic_soft_measures}. Usually, these policies do not imply any  physical change of the infrastructure. In fact, they rely on economic incentives or leverage psychological phenomena to change the drivers' habits. Most of the solutions based on this approach  lack strong theoretical fundations, and an  \textit{a posteriori} analysis is performed to study their consequences. In \cite{riggs:2017:painting_fence}, the author explores the effectiveness of: monetary incentives, gifts and social nudges tapping into altruistic values.  
 A personalized set of incentives (mostly monetary) is proposed in \cite{hu:2015:incentive_based_ADM}, where a platform is introduced that enables the commuters to receive incentives if they change their departure time to off-peak hours or use an alternative. Several other pilot studies have been  performed and they have experimentally validated the benefits of soft policies, see \cite{benalia:2011:changing_commuters_bahvior} among others. 
It is important to stress that these two classes of measures  are not always mutually exclusive but they can be used in combination to amplify the final effect of congestion alleviation, as we advocate in this paper. 
\subsection{Smart charging of \glspl{PEV}}
The continuous growth of the number of \glspl{PEV} is also due to the improvements in the smart charging, allowing the vehicles to charge up to $150\,\tup{kW}$. This technology increases the appeal of short stops for the users, making the \glspl{PEV} more similar to fuel vehicles. 
This motivates several studies on how the \gls{PEV} drivers may optimize their charging schedules and how they affect the distribution network. Some classic results~\cite{cenedese:2019:PEV_MIG,J_Hiskens_2013}, tackle the problem of high peaks in the energy demand by proposing a dynamic energy price that leads to a change in the  charging habits of the  \gls{PEV} owners, and consequently  to the so-called ``valley filling'' effect~\cite{parise:colombino:2014:mean_field_charging_PEV}. Some recent works considered smart charging coupled with mobility. In \cite{xu:2018:EV_urban_mobility_nature}, the smart charging problem is enhanced by considering also the travel habits. However, the goal is solely to decrease the energy peak demand rather than the traffic congestion level.
Other works focus on optimizing the charging of the \glspl{PEV} to decrease their travel time~\cite{wentao:2018:congestion_patterns_EV,razo:2016:smart_charging_for_highway_travel}. However, in these works the overall congestion level is not taken into account in the decision process. 
For this reason, we cannot consider these solutions as a form of \gls{ATDM}. To the best of the authors' knowledge, it is still an open, yet appealing, problem to develop \gls{ATDM} strategies based on smart-charging of \glspl{PEV} whose main goal is  traffic congestion alleviation.
\subsection{Paper contribution}
Inspired by the conventional ramp metering control and motivated by the rising number of \glspl{PEV}, we propose for the first time a novel \gls{ATDM} based on soft measures (via monetary incentives) that leverages smart fast charging of \glspl{PEV} in the road to alleviate traffic congestion during rush hours. Specifically, we propose a dynamic energy price discounted proportionally to the (predicted) congestion level. This approach  encourages the \gls{PEV} owners to stop for charging when the congestion level is (going to be) high, thus aligning the goal of the traffic control with the drivers' self-interest. In the following, we emphasize our main contributions: 
\begin{itemize}
\item We use for the first time the electricity price as control input for the \gls{ATDM}. While historically, traffic control had suffered from a lack of control means, this additional control input may prove itself essential to achieve the desired results by acting in synergy with the classical \gls{TDM}.
%
%
\item We enrich the \gls{CTM} with the introduction of \gls{r2s} and \gls{s2r} flows, newly defined to model the entering and leaving of the \glspl{PEV} in and out of a \gls{CS}.
\item We carry out a formal analysis of the effects of the presented soft policy by describing the decision process of the \gls{PEV} drivers as a  generalized exact potential game.  
\item We propose a semi-decentralized control scheme ensuring that the \glspl{PEV} involved in the decision reach an optimal charging schedule that represents their individual best trade-off between monetary saving and travel time.
\end{itemize}
In the second part of this work  \cite{cenedese:2020:highway_control_pII}(Part II: Case Study), we validate this \gls{ATDM} strategy  on a simulated  highway stretch between The Hague and Rotterdam, in The Netherlands.
%


\section{Cell Transmission Model with Charging Station} 
\label{sec:CTM}
 We consider a freeway stretch without ramps and only one \gls{CS} where \glspl{PEV} may stop. 
In the literature, the most used  model for traffic control is the \gls{CTM}, see~\cite{daganzo:1995:CTM_part2}. 

\begin{figure}[t]
\centering
\includegraphics[trim= 12 30 222 14, clip,scale= 0.88]{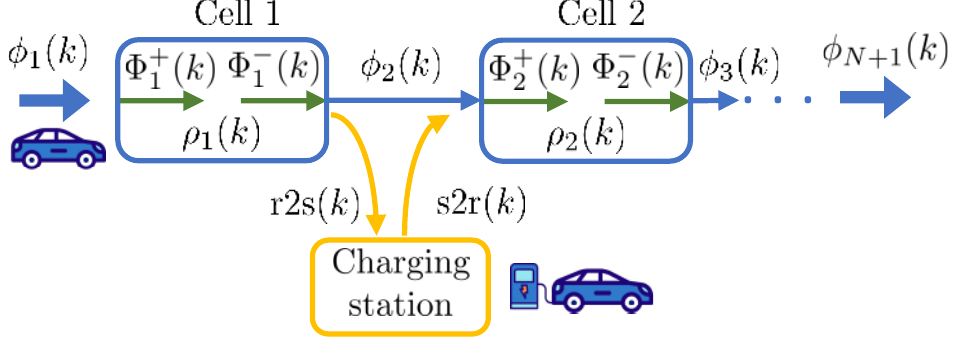}
\caption{Partitioning of the highway in cells and \gls{CS} for the \glspl{PEV}; compact graphical representation of the \gls{CTM} and the notation for the first two cells. }
\label{fig:ctm_model}
\end{figure}

Here, we explicitly introduce a revised version of the \gls{CTM} described in~\cite[Sec.~3.3.1]{ferrara2018freeway} adapted to  our problem. We consider the discretized version of the model where each time interval $[kT,(k+1)T)$ of length $T$ is denoted by an integer $k\in\N$.
 The highway stretch is modeled as a chain of $N$ subsequent \textit{cells} (Figure~\ref{fig:ctm_model}). The vehicles in each cell $\ell\in\mc N\coloneqq \{1,\dots,N\}$ are a mixture of \glspl{PEV} and non-\glspl{PEV} moving  at a constant speed. Two subsequent cells are connected via an \textit{interface} that models  a certain flow of vehicles, whose value depends on the cells' demand and supply capabilities. 
Without loss of generality, we assume that the \gls{CS} is located between the first two cells. 
To formalize the \gls{CTM}, for every cell $\ell\in\mc N$ and interval $k\in\N$, we introduce the following set of variables:
\begin{itemize}
\item $	\rho_\ell(k)\: [\textup{veh/km}]$: traffic density of  cell $\ell$ during  $k$; 
\item $\Phi_\ell^+(k) \:[\textup{veh/h}]$ (resp. $\Phi_\ell^-(k)$):  total flow entering (exiting) the cell $\ell$ during $k$;
\item $\phi_\ell(k)\:[\textup{veh/h}]$: flow entering cell $\ell$ from cell $\ell-1$ during $k$; $\phi_{1}(k)$ (resp. $\phi_{N+1}(k)$) is the flow entering (exiting) the highway during the same interval;
\end{itemize}
We enrich the conventional \gls{CTM} introducing two flows:
\begin{itemize}
\item ${\text{\gls{r2s}}}(k)\:[\textup{veh/h}]$: flow of \glspl{PEV} entering the \gls{CS} during $k$;
\item ${\text{\gls{s2r}}}(k)\:[\textup{veh/h}]$:  flow of \glspl{PEV} exiting the \gls{CS} during $k$. 
\end{itemize} 
Then, we associate a set of fixed parameter to each cell $\ell$:
\begin{itemize}
\item $L_\ell\:[\tup{km}]$: cell length;
\item $\overline v_\ell\:[\tup{km/h}]$: free-flow speed;
\item $w_\ell\:[\tup{km/h}]$: congestion wave speed;
\item $q_\ell^{\max}\:[\tup{veh/h}]$: maximum cell capacity;
\item $\rho_\ell^{\max}\:[\tup{veh/km}]$: maximum jam density.
\end{itemize}
Each cell can be seen as an input-output system where the inflow is the input and the outflow the output.
The dynamics of the density $\rho_\ell$ of cell $\ell\in\mc N$  read as
\smallskip
\begin{equation}
\label{eq:rho_dyn}
\rho_\ell(k+1) = \rho_\ell(k) + \dfrac{T}{L_\ell}\left( \Phi_\ell^+(k)-\Phi_\ell^-(k)\right) \:,
\end{equation}
where the inflow and outflow are defined as
\smallskip
\begin{subequations}
\label{eq:Phi_+_-}
\begin{align}
\label{eq:Phi_-}
\Phi^-_\ell(k) &\coloneqq \begin{cases}
 \phi_{\ell+1}(k) + {\text{\gls{r2s}}}(k) & \text{if } \ell=1 \\
 \phi_{\ell+1}(k) & \text{otherwise}
 \end{cases}\\ 
\Phi^+_\ell(k) &\coloneqq \begin{cases} \phi_\ell(k) + {\text{\gls{s2r}}}(k) & \quad \text{if } \ell=2 \\
\phi_\ell(k) & \quad\text{otherwise}.
\end{cases}
\end{align} 
\end{subequations}

Thus, the flows entering and exiting the \gls{CS} modify only the  definition of the in-flow  of cell $2$ and out-flow from cell $1$.

\smallskip
\begin{remark}[{\rm r2s} and {\rm s2r} flows]
The concepts of  \gls{r2s} and \gls{s2r} flows are inspired by the ``off-ramp'' and ``on-ramp'' flows~\cite{daganzo:1995:CTM_part2}, respectively, and used  to model the temporary stop of some \glspl{PEV} at the \gls{CS}, which leads, differently from the off- and on-ramp flows, to a mutual dependency between ${\text{\gls{r2s}}}$ and ${\text{\gls{s2r}}}$. We investigate this further in Sections~\ref{sec:decision_making_proc}~and~\ref{sec:sol_decision_proc}. In the literature, only the on-ramp flow can be controlled, e.g. via a toll, while our control action influences the off-ramp flow as well. \hfill\QEDopen
\end{remark}
\smallskip


The \textit{demand} $D_{\ell-1}(k)$ of cell $\ell-1$  and the \textit{supply} $S_\ell(k)$ of cell $\ell$ directly influence the admissible flow between the two cells. The former is the flow that cell $\ell-1$ can send to cell $\ell$ in the time interval $k$, while $S_\ell(k)$ describes the flow that cell $\ell$ can receive in the same interval:
\smallskip
\begin{subequations}
\label{eq:demand_supply}
\begin{align}
D_{\ell-1}(k) & \coloneqq \min\left\{ \overline v_{\ell-1}\rho_{\ell-1}(k)\, ,\: q_{\ell-1}^{\max} \right\} \:,\\
S_{\ell}(k) & \coloneqq \min\left\{ w_{\ell}\big(\rho_{\ell}^{\max}-\rho_{\ell}(k)\big)\, ,\: q_{\ell}^{\max} \right\} \;.
\end{align}
\end{subequations}  
The relations in \eqref{eq:demand_supply} directly define $\phi_\ell(k)$ in \eqref{eq:Phi_+_-}. In fact, if $\ell\in\{3,\,\dots,\, N\}$, then the flow between the cells reads as $\phi_\ell \coloneqq \min\left\{ D_{\ell-1}(k),\,S_\ell(k) \right\}$. On the other hand, the flow $\phi_2$ between cell $1$ and $2$ is described by a more complex relation, due to the presence of the \gls{CS}:  
\smallskip
\begin{equation}
\label{eq:phi_2}
\begin{cases}
\phi_2 \coloneqq D_1- {\text{\gls{r2s}}} & \text{if}\: D_1-{\text{\gls{r2s}}} \leq S_2-{\text{\gls{s2r}}}\\
\phi_2 \coloneqq S_2-{\text{\gls{s2r}}} & \text{otherwise}\:,
\end{cases}
\end{equation}
where the time dependency is omitted. The first case in \eqref{eq:phi_2} reflects the free-flow scenario, while the second reflects the presence of a congestion, as the supply of cell $2$ is saturated by $\phi_2(k)$ and ${\text{\gls{s2r}}}(k)$. Finally, $\phi_1(k)$ and $\phi_{N+1}(k)$ are the input and output flows of the \gls{CTM}, respectively.  

Throughout this section, we have defined the whole \gls{CTM}  except for \gls{r2s}$(k)$ and \gls{s2r}$(k)$. The remainder of the paper is devoted to design the decision process that the \glspl{PEV} carry out to choose whether or not it is worth    stopping at the \gls{CS}. In turn, this determines \gls{r2s}$(k)$ and \gls{s2r}$(k)$, as we show in Section~\ref{sec:sol_decision_proc}. 
\section{Decision making process}
\label{sec:decision_making_proc}
We assume the presence of a \gls{HO} that aims at minimizing the congestion. 
It does that by discounting the energy price if the level of congestion grows (or is expected to grow). 
In this setup, the \gls{HO} would have the role of the  so-called \textit{choice architect}, by designing the price at all  time intervals. Our solution leverages two main aspects:  first, if the road is congested, the benefit of keep driving decreases, due to a longer travel time, and, at the same time, stopping at a \gls{CS} to charge becomes more profitable due to an energy price discount. Secondly, we take advantage of  the \textit{range anxiety}, which  is a well-known cognitive bias affecting \gls{PEV} drivers making them impatient to stop at a \gls{CS} even if they do not strictly need it~\cite{nuubauer:2014:range_anxiety}. 
 
We model the multi-agent decision process of   the \glspl{PEV} exiting cell $1$, at every time interval $k$, by defining  a set of interdependent local optimization problems. Each \gls{PEV} (or agent/player) aims at minimizing its own local cost function subject to a set of constraints, where couplings between the agents arise in both the cost functions and the constraints. From a mathematical point of view, specific in our problem setup, the collection of all these optimization problems determines a \textit{mixed-integer potential game} subject to best-response dynamics. The output of the decision making process (or game) is the set of all the choices (or strategies) of the \glspl{PEV} to stop or not  at the \gls{CS}, which affects the \gls{r2s} flow. The \gls{s2r} flow is instead a consequence of how long the agents decide to linger at the \gls{CS}. 
\subsection{Cost function}
Next, we design the cost function of the \glspl{PEV} exiting cell $1$ that are involved in the decision making process.
We postulate that the interest of each driver is twofold:  he is interested in minimizing the travel time, while he is also willing to save money for charging his  \gls{PEV}. Between the two, in most cases, the primary concern will be the travel time, especially in  normal traffic conditions, when  no discount is present. Nevertheless, the travel time aspect becomes less relevant if a heavy congestion arises; in this situation, the relative impact of the travel time spent at the \gls{CS} decreases, and at the same time, the discounted energy price may steer the decision of the agent  towards the choice of stopping to charge the \gls{PEV}. 

\subsubsection{Number of vehicles}
At each time interval $k\in\N$, the number of vehicles involved in the decision process is $n_{\tup{EV}}(k)$, which may vary due to the traffic conditions. From~\eqref{eq:Phi_+_-}, we show that the total number of vehicles exiting cell $1$  during $k$ is $\Phi_1^-(k) \: T$. 
Between those,  the fraction of \glspl{PEV} is denoted by $p_{\tup{EV}}\in [0,1]$. By relying on \eqref{eq:Phi_-} and \eqref{eq:phi_2}, we attain
\begin{equation}
\label{eq:n_EV}
\begin{cases}
n_{\tup{EV}} = \lfloor p_{\tup{EV}}D_1 \,T\rfloor   \,,\quad \text{if}\: D_1 - {\text{\gls{r2s}}} \leq S_2- {\text{\gls{s2r}}}\\
n_{\tup{EV}} = \lfloor p_{\tup{EV}}( S_2-{\text{\gls{s2r}}} + {\text{\gls{r2s}}} )T\rfloor \,,  \quad \text{otherwise}
\end{cases}
\end{equation}
where the time dependency is omitted and $\lfloor x \rfloor\in\N$ denotes the floor of $x\in\R$. 

To compute $n_{\tup{EV}}(k)$ via~\eqref{eq:n_EV}, the value of \gls{r2s}$(k)$ is necessary, even though it is  the solution of the decision process that we are defining. Thus, it is not possible to exactly define $n_{\tup{EV}}(k)$. At the same time, \gls{s2r}$(k)$ does not affect the computation of $n_{\tup{EV}}(k)$, since it is due to \glspl{PEV} already at the \gls{CS}, so not involved in the decision process arising at time  $k$. To overcome this impasse, we compute the number of agents $n(k)$ involved in the game under the assumption of maximum congestion, namely if no agent stops at the charging station (\gls{r2s}$(k)=0$).
 Then the number of agents taking part in the decision process is obtained as
\smallskip
\begin{equation*}
\label{eq:n_EV_s0}
\begin{cases}
n(k) = \lfloor p_{\tup{EV}} \:D_1(k)T \rfloor\:,\quad\text{if}\: D_1(k) + {\text{\gls{s2r}}}(k) \leq S_2(k)\\
n(k) = \lfloor p_{\tup{EV}} (S_2(k) -{{\text{\gls{s2r}}}}(k))T \rfloor \:,\quad \text{otherwise}\:.
\end{cases}
\end{equation*}
This assumption not only implies that $n(k)$ can be computed for every $k$, but also that all the vehicles involved in the game manage to exit cell $1$ during the time interval $k$, and therefore being able to implement their strategies.

\subsubsection{Decision variables and the SoC dynamics}
The time-varying set indexing the $n(k)$ \glspl{PEV} taking part in the game during the  $k$-th time interval is denoted  by  $\mc I(k)\coloneqq\{1,\dots,n(k)\}$. The decisions of the agents are performed over a prediction horizon $\mc T(k)$ of $T_{\tup h}$ time intervals. The length of the time intervals in the decision making process may be longer than the one used in the \gls{CTM}. Specifically, we assume intervals of length $lT$, with $l\in\N$. Thus, the \glspl{PEV} should plan their behavior over the set of intervals $\mc T(k)\coloneqq\{k,k+1, \dots,k+T_{\tup h}\}$, where each index denotes an interval of length $lT$, e.g.,  $k\in\mc T(k)$ represents here $[kT,kT+lT)$ and similarly $k+1\in\mc T(k)$ denotes $[kT+lT,kT+2lT)$.  

The \gls{SoC} of the battery of every \gls{PEV} $i\in\mc I(k)$ at time $t\in\mc T(k)$ is denoted by $x_i(t)\in [0,1]$, where $x_i(t)=1$ represents a fully charged battery, while $x_i(t)=0$ a completely discharged one.  
The amount of energy purchased by agent $i\in \mc I(k)$ during the time period $t\in\mc T(k)$ is $u_i(t)\geq 0$. For every interval $t\in\mc T(k)$, the \gls{SoC} reads as 
\begin{equation} \label{eq:SoC_dyn}
x_i(t+1) = x_i(t) + b_i u_i(t)\:,
\end{equation}  
where $b_i=\frac{\eta_i}{C_{i}}>0$ is a coefficient associated to the battery efficiency $\eta_i \in (0,1]$ and  capacity $C_{i} > 0$. Let us introduce the variable $\delta_i(t)\in\B\coloneqq\{0,1\}$ as a binary decision variable, which takes value $\delta_i(t)=1$ if the vehicle is actively charging at the \gls{CS} and $\delta_i(t)=0$ otherwise: 
\begin{equation}
\label{eq:logical_delta_u}
[\delta_{i}(t) = 1] \iff [u_i(t)> 0],\quad \forall t\in\mc T(k),\, \forall i\in\mc I(k).
\end{equation}
The logical implication above entails that the energy purchased by \gls{PEV} $i$ is positive if and only if $\delta_i(t)=1$, therefore we define $
	u_i(t)\in \mc U(t)\coloneqq[\underline{u}\delta_i(t),\overline{u}\delta_i(t)]$, with $\overline{u}>0$ (respectively $\underline{u}>0$) being the maximum (minimum) energy that the vehicle can receive from the \gls{CS} in a single time interval.  
We define a first collection of decision variables associated with each \gls{PEV}  $i\in\mc I(k)$, over the whole horizon $\mc T(k)$, as the collective vectors $u_i \coloneqq \col( (u_{i}(t))_{t\in\mc{T}})$, $\delta_i \coloneqq \col((\delta_i(t))_{t\in\mc{T}})$ and the evolution of the \gls{SoC}, $x_i \coloneqq \col((x_i(t))_{t\in\mc{T}})$.

\subsubsection{Travel time and congestion}
\label{sec:travel_time}

An important quantity influencing the \gls{PEV} decisions is the additional time $\xi(t)\geq 0$  that a \gls{PEV} would experience due to the presence of congestion. Specifically, $\xi(t)$ denotes the difference between the travel time that a \gls{PEV} experiences to actually  travel throughout the cells $\{2,\dots ,N\}$ and the one it would spend in conditions of free flow.    
It provides insightful information on the traffic evolution, allowing the \glspl{PEV} to discern whether or not they prefer to stop at the \gls{CS}. 
If agent $i\in\mc I(k)$ decides to stop for charging, the congestion it will experience, when it merges back in the mainstream during the time interval $ t_i\in\mc T(k)$, depends also on those that were behind it at the time of the decision $k< t_i$. 
Among all the vehicles exiting cell $1$ in the time interval $(k, t_i ]$, the \glspl{PEV} have the possibility to stop  for charging, deciding via a process akin to the one that agent $i$ is  currently carrying out.  For this reason, the exact value of $\xi(t)$ cannot be computed in advance by agent $i$  for the whole prediction horizon. We work around this difficulty by adopting a conservative approximation of $\xi(t)$,  computed assuming that  all the \glspl{PEV}  in $\mc I(k)$ and the ones following them  do not stop at the \gls{CS}. This leads to a value of $\xi(t)$ that over-estimates the  actual  experienced travel time. This approximation allows the agents to  cope with the worst-case scenario, hence being able to meet possible time constraints. Moreover, it can be computed at every time interval and provides insightful information on the potential traffic evolution.


For a cell $\ell\in\mc N $, the vehicles' speed  is attained as   $v_\ell(k)=\Phi_\ell^-(k)/\rho_\ell(k)$. 
If an agent $i$ enters cell $\ell$ during the time interval $k$ and it takes $\overline t\in\N$ intervals to travel through it, then the velocity at which it will move when it enters the next cell is  $v_{\ell+1}(k+\overline{t})$. This observation motivates the following recursive, but implementable, definition 
\smallskip
\begin{equation}
\label{eq:xi_cell}
\quad\xi(t)\coloneqq  \xi_N(t) \:,\: \forall t\in\mc T(k)
\end{equation} 
where
\begin{align*}
\begin{cases}
\xi_1(t) = 0 \\
\xi_\ell(t) = \xi_{\ell-1}(t) + \dfrac{L_\ell}{\hat v_\ell(t+\xi_{\ell-1}(t))} - \dfrac{L_\ell}{\overline v_\ell} \,, \: \forall\ell\{2,\dots,N\}.
\end{cases}
\end{align*}
Here, $\hat v_\ell$ denotes the vehicles' speed computed in the worst-case scenario described above. The value $L_\ell/\overline v_\ell$ represents the travel time in the case of no congestion in the cell $\ell$. Under the assumption above, $\xi_\ell(t)$ can be always  computed by letting the \gls{CTM} evolve freely. It is worth noticing that, if the \glspl{PEV} will experience no congestion along the whole freeway, i.e., $\hat v_\ell(t)=\overline v_\ell$ for all $\ell\in\mc N$, then  $\xi(t)=0$. 
 

Another important quantity related to the congestion is the number of vehicles that leave the \gls{CS} at every time instant. In fact, if an agent leaves the \gls{CS} when many others are also merging back into the mainstream, 
it may experience high levels of congestion. To model this phenomenon, we introduce a binary variable $\vartheta_i\in\B$ for every $i\in\mc I(k)$. 

\begin{figure}[t]
\centering
\includegraphics[trim= 0 0 0 20, clip,width=.5\textwidth]{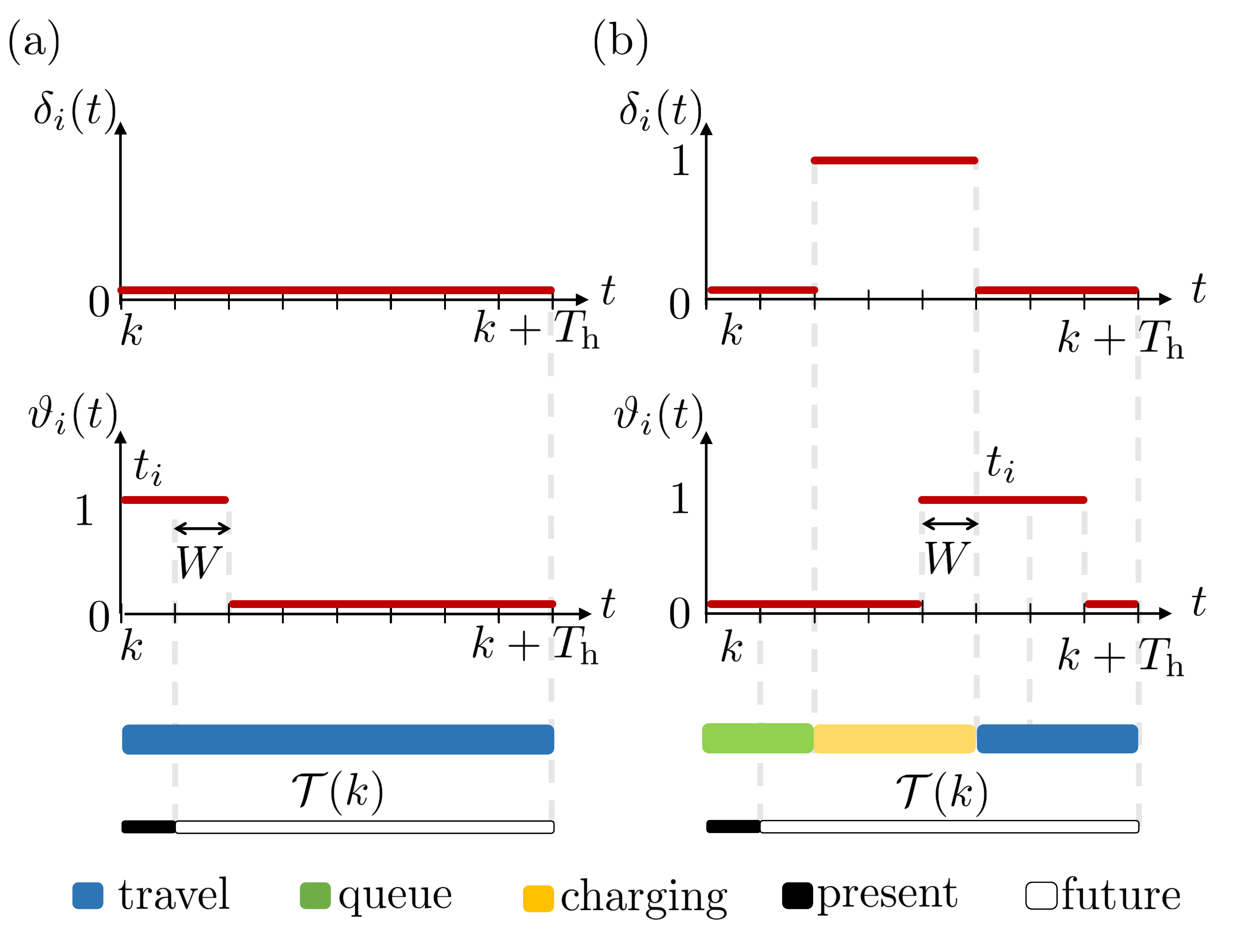}
\caption{Feasible choices of $\delta_i$ and $\vartheta_i$  for $i\in\mc I$  when $l=1$ and $W=1$. The two illustrations show when: (a) the \gls{PEV} does not stop at the \gls{CS}; (b) the \gls{PEV} decides to stop.}
\label{fig:delta_theta}
\end{figure}

For every \gls{PEV}  $i\in \mc I(k)$ entering cell $2$ at time $ t_i \in\mc T(k)$, and for every $t\in\mc T(k)$, we define     $\vartheta_i(t)=1$ if $ t\in \{t_i-W, \dots, t_i+W\}$, and $0$ otherwise. Thus, $\vartheta_i$ is a rectangular function of width at most $2W+1$ intervals and at least $W+1$ (Figure~\ref{fig:delta_theta}). This variable is used to capture the influence of the \glspl{PEV} entering cell $2$ around the same time  as agent $i$. The value of $W\in\N$ depends on $lT$. In fact, if $lT$ is large, then \gls{PEV} $i$ will not experience the congestion due to the \glspl{PEV} that precede or follow him, so $W=0$. On the other hand, if $lT$ is small, the value of $W$ has to be high to model correctly the possible congestion due to those agents that enter cell $2$ around the same time as agent $i$. We elaborate further on this in the next section.
Also in this case, we denote the collective vector over the whole $\mc T(k)$ as $\vartheta_i \coloneqq \col((\vartheta_i(t))_{t\in\mc T(k)})$. 
Therefore, for every $t\in\mc T(k)$ and $i\in\mc I(k)$, $\xi_{i}^{\tup{CS}}(t)$ approximates the extra time that  agent $i$ would experience due to those \glspl{PEV} entering cell $2$ around time $t$ and it reads as
\begin{equation}
\label{eq:xi_game}
\xi_i^{\tup{CS}}(t) \coloneqq \gamma \left(\sum_{\bar k <k}\sum_{j\in \mc I(\bar k)} \vartheta_j(t) + \sum_{j\in\mc I(k)\setminus\{i\}}  \vartheta_j(t)\right)\:.
\end{equation}
The first double summation, denoted by $\vartheta^{\tup{old}}(t)$ for all $t\in\mc T(k)$, represents the number of \glspl{PEV}, that already completed the decision process, entering cell $2$ during one of the intervals $\{\max(k,t-W),\dots,t+W\}$. The coefficient $\gamma>0$ is proportional to the average amount of time agent $i$ spends for every \gls{PEV} entering cell $2$ during the intervals $\{\max(k,t-W),\dots,t+W\}$. This coefficient may be estimated via historical data and engineering understanding or based on the worst-case scenario.


\subsubsection{Energy price} 
In our model the dynamic energy price $p(k)$  is discounted  by the \gls{HO} in conjunction with a traffic congestion. 
 On the other hand, it is also linked to the local energy demand required in the distribution network, i.e., $d(k) + u^{\textup{PEV}}(k)$, where $u^{\textup{PEV}}(k)\coloneqq \sum_{\bar k \in\N}\sum_{i\in\mc I(\bar k)} u_i(k)$ is the total energy purchased by the \glspl{PEV}  and $d(k)$ denotes the base energy demand of the local network. During the time interval $k$, we can study the congestion by looking at how much the travel time increases w.r.t. the free-flow case, for each cell $\ell\in \mc N$. This quantity is defined by $\Delta_\ell(k)\coloneqq \frac{L_\ell}{v_\ell(k)} - \frac{L_\ell}{\overline v_\ell}$, and $\Delta_\ell(k)\geq 0$. Here we assume it is used by the \gls{HO} to link the price to the congestion level, namely the higher $\Delta_\ell(k)$ the lower the price.
Thus, the energy price that the \gls{HO} imposes for every unit of energy purchased reads as 
\smallskip
\begin{equation}
\label{eq:price}
p(k) \coloneqq c_1 \: d(k) + c_2  u^{\textup{PEV}}(k) - c_3\sum_{\ell=2}^N \Delta_\ell(k)  \:,
\end{equation}
where $c_1,c_2, c_3>0$ are scaling parameters tuned by the \gls{HO}.    

We note that the exact energy price applied in the future time intervals $t\in \mc T(k)$ cannot be computed in advance by the \gls{HO}, since it depends on the traffic evolution, which is not completely known due to the arbitrary future choices of the drivers. Nevertheless, to allow the \glspl{PEV} to perform an informed choice, we let the \gls{HO} compute an estimation $\hat p(t)$ of the real $p(t)$ for the whole prediction horizon $ \mc T(k)$. Then, this value is broadcast to the \glspl{PEV} in $\mc I(k)$ and it is used by them to execute the decision process. If the congestion grows, then the price should drop, even though, intuitively, the discounted price leads to a larger number of \glspl{PEV} stopping, and consequently an increment of $u^{\textup{PEV}}(k)$. 
We define $\hat \Delta_\ell(k)$ as an approximate value of $ \Delta_\ell(k)$, which is computed by assuming that no agent exiting cell $1$ during the prediction horizon $\mc T(k)$ stops at the \gls{CS}. This assumption translates into \gls{r2s} $(t)=0$ for all $t\in\mc T(k)$. 
  The density and the flow during the time interval $t\in\mc T(k)$ are attained by letting the \gls{CTM} evolve freely. Therefore, the approximation of the additional time spent by the agent, due to the congestion in the cell $\ell$,  reads as    
\smallskip
\begin{equation}
\label{eq:delta_hat}
\hat \Delta_\ell(t)\coloneqq  \frac{L_\ell}{\hat v_\ell(t)}-\frac{L_\ell}{\overline v_\ell}\:.
\end{equation}

The value of $\hat \Delta_\ell(t)$  overestimates the additional travel time spent due to the congestion level on the road during the interval $t$. 
We introduce two time-varying vectors of offsets and coefficients, $\boldsymbol \beta_0(k) \coloneqq \col ((\beta_0(t))_{t\in\mc T(k)})$ and $\boldsymbol \beta_1(k) \coloneqq \col ((\beta_1(t))_{t\in\mc T(k)})$ respectively, and define the estimated price, for every time interval $t\in\mc T(k)$, by 
\smallskip
\begin{equation}
\label{eq:price_hat}
\hat p(t) \coloneqq c_1 \: d(t) - \left[\beta_0(t) +\beta_1(t)  \sum_{\ell=2}^N \hat \Delta_\ell(t)  \right]\:.
\end{equation}

At every time instant $k\in\N$, the \gls{HO} may use historical data on the traffic flow to compute the  values of $\boldsymbol \beta_0(k)$ and $\boldsymbol \beta_1(k)$ that  are supposed to minimize the error between the real and estimated price. This may be done with several techniques (e.g. linear regression or Bayesian estimation). 
\smallskip
\begin{remark}
The definition in \eqref{eq:price_hat} implies that the estimated price is not affected by the strategies of the other agents in the game. Nevertheless, the strategies implemented by the \glspl{PEV} involved during  $k$, directly influence the estimated price used by the \glspl{PEV} that will play the game during $k+1$. Therefore, the price dynamically changes over time and is assumed to be fixed only inside the single decision process.\hfill$\square$ 
\end{remark}



\subsubsection{Cost function formulation}
The goal of each \gls{PEV} is  to find the best trade-off between saving money, and  travel time. These two cost terms are described for every \gls{PEV} $i\in\mc I(k)$ by the functions $J^{\tup{price}}_{i}$ and $J^{\tup{time}}_{i}$, respectively.
The amount agent $i$ saves by charging at a discounted price depends on the total energy it  purchases:
\begin{equation*}
\label{eq:J_price}
J^{\tup{price}}_{i}(u_i|k) \coloneqq \sum_{t\in\mc T(k)} (\hat p(t)-\bar p_i)u_i(t)\:,
\end{equation*}
where $ \bar p_i>0$ represents the average cost agent $i$ would experience via standard fast charging, and it might vary between \glspl{PEV}. 
Next, we define the cost associated to the total travel time experienced by vehicle $i$ by
\smallskip
\begin{equation*}
\label{eq:J_time}
J^{\tup{time}}_{i}(\vartheta_i,\bs{\vartheta}_{-i}|k) \coloneqq \sum_{t\in\mc T(k)} \chi(t) \big[(t-k)\upsilon +
 \xi(t)+\xi^{\tup{CS}}_i(t)\big]\vartheta_i(t) ,
\end{equation*} 
where the notation $\bs{\vartheta}_{-i}$ is  used to denote  $ \col((\vartheta_j)_{j\in\mc I(k)\setminus \{i\}})$. The quantity $(t-k)$ weights the time spent at the charging station, while the rest approximates  the additional time spent if the agent enters cell $2$ during the time interval $t$. The parameter $\upsilon>0$ weights the different perception that the agent has in spending time at the \gls{CS} or in a congestion. The time-varying factor $\chi(t)$ normalizes the cost function with respect to the width of the rectangular function $\vartheta_i$.   
Note that the presence of $\xi_i^{\tup{CS}}(t)$ creates a coupling between all the decisions of the \glspl{PEV} in the game. In $J^{\tup{time}}_{i}$, the presence of $\vartheta_i$ entails that only some elements of the summation are not zero. Furthermore, its rectangular shape implies that the decision of agent $i$ depends also on those agents that enter cell $2$ during an interval distant at most $2W$ intervals from the one in which $i$ will enter cell $2$, see Figure~\ref{fig:delta_theta}. As anticipated, this feature models the different  speeds of the vehicles in a cell. 


Each agent may weight differently the two objectives thus we model the final cost as a convex combination of the two:
\smallskip
\begin{equation}
\label{eq:J_i}
J_i(u_i,\vartheta_i,\bs{\vartheta}_{-i}|k) \coloneqq  \alpha_iJ^{\tup{price}}_{i}+ (1-\alpha_i)J^{\tup{time}}_{i}\,,
\end{equation}
 for some $\alpha_i\in(0,1)$. We study the effects of this parameter on the performance  in \cite{cenedese:2020:highway_control_pII}(Part II: Case study).    Finally, we highlight that the nature of the approximation of $\xi^{\tup{CS}}_i$ and the estimation $\hat p$ are intrinsically different. In fact, despite both uncertainties are due to the presence of the human in the loop, the second is part of the policy designed by the \gls{HO} to reduce the congestion, while the first is used to model the drivers' local decision. Consequently, the complete policy includes the actual price applied and its estimation over the prediction horizon, that are broadcast by the \gls{HO} to the \glspl{PEV} and  used to influence their decision.  
\subsection{Local and coupling constraints}
\label{sec:local_coup_constraints} 
We model the constraints on the drivers' possible choices as  a collection of logical implications.
First, we impose that the $2W+1$ intervals in which $\vartheta_i=1$ are consecutive. To do so, we require that, for all $i\in\mc I(k)$, $\vartheta_i$ changes its value from $0$ to $1$ and back to $0$ only once:
\smallskip
\begin{subequations}
\label{eq:varth_consecutive}
\begin{equation}
\sum_{p \in\mc T(k)} (1-\vartheta_i(p-1))\vartheta_i(p) = 1 
\end{equation}
\begin{equation}
\sum_{p \in\mc T(k)} \vartheta_i(p-1) (1-\vartheta_i(p))  = 1 \:, 
\end{equation}
\end{subequations}
where $\vartheta_i(k-1)=0$, so the raising edge must precede the falling one. Then, we force the intervals in which $\vartheta_i(k)=1$ to be consecutive, and hence for all $i\in \mc I(k)$ and $t \in \mc T(k)$ it must hold that
\smallskip
\begin{equation}
\label{eq:consecutive_intervals}
\begin{split}
\vartheta_i(t) (1-\vartheta_i&(t+1))\bigg( \sum_{p\in \mc T(k)}  \vartheta_i(p)-\mathfrak{P} \bigg)=0\\ 
& \mathfrak{P} \coloneqq \min\{t-k+W+1, 2W+1\}\:.
\end{split}
\end{equation}


Clearly, if a \gls{PEV} enters cell $2$ at time $t_i$, it cannot charge in the remaining time intervals (Figure~\ref{fig:delta_theta}), and thus we obtain the following relation, for all $t\in\mc T(k)$ and $i\in\mc I(k)$: 
\smallskip
\begin{equation}
\label{eq:cnd_no_stop}
\begin{split}
\left[ \vartheta_i(t-1)(1- \right. & \left. \vartheta_i(t)) = 1 \right] \implies [\delta_i(r) = 0]\:,\,\\
&  \forall r\in\{t-W-1,\dots,k+T_{\tup h}\}\:.
\end{split}
\end{equation}
This condition models also the case in which a \gls{PEV} is not stopping, so $\vartheta_i(k)=\dots=\vartheta_i(k+W)=1$ and \eqref{eq:cnd_no_stop} implies that the \gls{PEV} is never charging, i.e., $\delta_i(t)=0$ for all $t\in\mc T(k)$. Furthermore, \eqref{eq:varth_consecutive}~and~\eqref{eq:consecutive_intervals} imply that the \glspl{PEV} disconnect at least $W+1$ time intervals before the end of the prediction horizon. Next, we impose that, when an agent is done with charging, it exits the \gls{CS}, and hence, for all $t\in\mc T(k)$, agent $i\in\mc I(k)$ has to satisfy
\begin{equation}
\label{eq:exit_CS}
\begin{split}
\left[ \delta_i(t-1)=1 \right] & \wedge \left[ \delta_i(t)=0  \right] \implies  [\vartheta_i(r) = 1 ]\\
  \forall r&\in\{\max(k,t-W),\dots,t+W\}\:.
\end{split}
\end{equation}
We impose that each \gls{PEV} charges for at least $\overline h\in\N$ consecutive time intervals. In fact, the value $lT$ may be small and it is unreasonable to allow a \gls{PEV}  to stop for charging for only one time interval (e.g. $2$ minutes). This translates into
\smallskip
\begin{equation}
\label{eq:cnd_charge4h}
[ \delta_i(t-1)=0 ] \wedge [ \delta_i(t)=1  ] \Rightarrow [\delta_i(t+h) = 1,\:\forall h \leq \overline h]\: .
\end{equation}  
Similarly, if a \gls{PEV} decides to stop, then  we assume it remains at the \gls{CS} for at least $2W+1$ time intervals, and hence
\smallskip
\begin{equation}
\label{eq:min_time_CS}
\sum_{t\in\mc T(k)} \vartheta_i(t) > W+1 \Leftrightarrow \vartheta_i(k)=\dots =\vartheta_i(k+W) = 0\,.
\end{equation}  

For each \gls{PEV}, the minimum level of \gls{SoC} necessary to reach the final destination from cell $2$ is denoted by $x_i^{\tup{ref}} \in (0,1]$. Thus, we assume that a \gls{PEV} can enter cell $2$ if $x_i(t)>x_i^{\tup{ref}}$, otherwise it must stop (or remain) at the \gls{CS} for charging, so for all $t\in\mc T(k)$,  
\smallskip
\begin{equation}
\label{eq:min_charge}
[x_i(t) < x_i^{\tup{ref}}] \implies [\vartheta_i(\max(k,t-W))=0]\:,
\end{equation}
where $\vartheta_i(\max(k,t-W))=0$ implies that \gls{PEV} $i$ cannot leave the charging station during the time interval $t$.
The next constraint limits the maximum amount of energy that the \gls{CS} can supply during each time interval by $u^{\tup{max}}>0$. Thus, for all $t\in\mc T(k)$, we have  the following coupling constraint on the connected \glspl{PEV}:  
\smallskip
\begin{equation}
\label{eq:max_energy_supply}
u^{\tup{old}}(t)+\sum_{i\in\mc I(k)} u_i(t) \leq u^{\tup{max}}\:,
\end{equation}
where $u^{\tup{old}}(t)\coloneqq \sum_{\bar k<k}\sum_{j\in\mc I(\bar k)} u_j(t)$ is the total energy that the agents, that already completed the decision process, planned to purchase  during the time interval $t$. 

Finally, we consider that if several \glspl{PEV} stop at the \gls{CS} simultaneously there can be a scarcity of charging plugs. Let $\bar \delta$ denote the total number of  plugs at the \gls{CS}. Then, we have
\smallskip
\begin{equation}
\label{eq:numb_plugs}
\delta^{\tup{old}}(t) - \sum_{i\in\mc I(k)} \delta_i(t) +  \leq \overline \delta \:, \:\forall t\in\mc T(k)
\end{equation} 
where $\delta^{\tup{old}}(t)$ is defined analogously to $u^{\tup{old}}(t)$.

The above constraints allow the \glspl{PEV} to stop at the \gls{CS} and do not start charging immediately (for example due to a lack of free plugs), and this may lead to the formation of a queue. We model the queue as a \textit{First-In-First-Out} (FIFO), i.e., the vehicles already waiting have the priority over the \glspl{PEV} entering it afterwards. This aspect is important to realistically model the \gls{PEV} behaviors, which would be hard to formalize without the use of mixed-integer variables. 

In Figure~\ref{fig:delta_theta}, we qualitatively represent a feasible choice of $\delta_i$ and $\vartheta_i$ for \gls{PEV} $i$ and how it is reflected  in the driver's behavior. 
In Figure~\ref{fig:delta_theta}a, agent $i$ does not stop at the \gls{CS}. In comparison, in Figure~\ref{fig:delta_theta}b,  the \gls{PEV} enters the \gls{CS}, but, since all the plugs are busy, it waits for the first two time intervals before connecting to the \gls{CS}. Once it finishes the charging phase, i.e., $\delta_i(t)=0$, it  merges back into the mainstream, according to~\eqref{eq:exit_CS}.

We conclude this section by introducing a preliminary formulation of the set of inter-dependent mixed-integer optimization problems that model  the decision process performed by the $n(k)$ \glspl{PEV} during every time interval $k\in\N$:
\begin{equation}\label{eq:MPC_first}
\forall i \in \mc{I}(k) : \left\{\begin{aligned}
&\underset{u_i, x_i, \delta_i, \vartheta_{i}}{\textrm{min}} & & J_i(u_i,\vartheta_i,\bs{\vartheta}_{-i}|\,k)\\
&\hspace{.5cm}\textrm{s.t.} & & \eqref{eq:SoC_dyn}, x_i(t) \in [0, 1], \delta_i(t)\in \B,\\
&&& u_i(t) \in \mc{U}(t), \vartheta_i(t)\in \B,\\
&&& \eqref{eq:logical_delta_u},\eqref{eq:varth_consecutive} \text{--} \eqref{eq:numb_plugs}, \; \forall t \in \mc{T}(k).
\end{aligned}\right.\tag{$\mc{P}$}
\end{equation} 

Several constraints in \eqref{eq:MPC_first} are expressed via logical implications, thus this problem should be mathematically reformulated to be solved. Specifically, in the Appendix we adopt a process akin to the one used in~\cite{cenedese:2019:PEV_MIG,fabiani2018mvad} to transform the logical implications into mixed-integer affine coupling constraints by additional auxiliary variables.

\section{Formulation of the mixed-integer game}


As a result of translating the logical implications into affine constraints, we recast \eqref{eq:MPC_first} as the following mixed-integer aggregative game, subject only to linear mixed-integer inequalities: 
\smallskip
\begin{equation}\label{eq:MPC_second}
\forall i \in \mc{I}(k) : \left\{\begin{aligned}
&\underset{ u_i, \dots,\nu_i}{\textrm{min}} & & J_i(u_i,\vartheta_i,\vartheta_i,\bs{\vartheta}_{-i}|\,k)\\
&\hspace{.5cm}\textrm{s.t.} & &  x_i(t) \in [0, 1], u_i(t) \in \mc{U}(t),\\
&&&  \delta_i(t),\vartheta_i(t),\psi_i(t)\in \B\\
&&&  \sigma_i(t),\omega_i(t)\in \B\\
&&& \varphi_i^{\tup{LH}}(t),\varphi_i^{\tup{HL}}(t), \nu_i\in \B\\
&&&  \mu_i^{(h)}(t)\in \B\,\: \forall h\leq \overline{h}\\
&&& g_i(t),q_i(t) \in\N\\
&&& \eqref{eq:max_energy_supply}\text{--}\eqref{eq:log2lin_13},
 \; \forall t \in \mc{T}(k).
\end{aligned}\right.\tag{$\mc{G}(k)$}
\end{equation}
The vector of all the decision variables in~\eqref{eq:MPC_second} is defined as
\begin{align*}
z_i\coloneqq \col (u_i,x_i,&\delta_i,\vartheta_i,\psi_i,\sigma_i,\varpi_i, \varphi_i^{\tup{LH}},\\
&\varphi_i^{\tup{HL}},\mu_i^{(1)} \dots,\mu_i^{(\overline h)},g_i,q_i,\nu_i)\in\R^{n_i}\:,
\end{align*}
and $\bs z \coloneqq \col((z_i)_{i\in\mc I(k)})$, we obtain a compact form of \ref{eq:MPC_second}:
\smallskip
\begin{equation}
\label{eq:game_compact}
\forall i \in \mc I(k) \::\quad \min_{z_i\in\mc Z_i(k)} J_i(z_i,\bs z_{-i}| k) \quad \text{s.t.} \: A\bs z\leq b 
\end{equation} 
where $\mc Z_i(k)$ is the set of strategy that satisfy the local constraints of $i$, while  $A$ and $b$ are of suitable dimensions and are used to describe all the coupling constraints between the agents. We denote the set of all feasible strategies of player $i\in\mc I(k)$ as 
\begin{equation}
\label{eq:Z_i}
\mc{Z}_{i}(\bs{z}_{-i}|k) \coloneqq \{ y \in \mc Z_i(k)\subset \R^{n_i} \mid A (y, \bs{z}_{-i}) \leq b \}\,,
\end{equation}
 where $(y, \bs{z}_{-i})$ indicates the collective strategy vector, with $y$ being any feasible strategy of $i\in\mc I(k)$. Then, the set of all the feasible collective strategies is $$\bs{\mc{Z}}(k) \coloneqq \left\{\bs{z} \in {\textstyle \prod_{i\in\mc I(k)} } \mc Z_i(k)\subset \R^{n} \mid A \bs{z} \leq b\right\},$$
  where $n\coloneqq {\scriptstyle\sum_{i\in\mc I(k)}} n_i$. 


Perhaps, the most popular notion of equilibrium for games like \ref{eq:MPC_second} is the Generalized Nash Equilibrium (GNE), where no agent can reduce its cost by unilaterally changing its strategy to another feasible one~\cite{cenedese_et_al:2019:TAC:proximal_point,cenedese:2019:ECC}. Here, we are interested in an approximate solution for mixed-integer games, i.e., \gls{MINE}.
\begin{definition}[$\varepsilon$-Mixed-Integer Nash equilibrium]\label{def:MINE}
	A set of strategies $\bs{z}^{*} \in \bs{\mc{Z}}$ is an \gls{MINE}, with $\varepsilon > 0$, of the game \ref{eq:MPC_second} if, for all $i \in \mc{I}$,
	\begin{equation*}
	J_i(z^{*}_i, \bs{z}^{*}_{-i}|k) \leq \underset{z_i \in \mc{Z}_i(\bs{z}^{*}_{-i}|k)}{\textrm{inf}} J_i(z_i, \bs{z}^{*}_{-i}|k) + \varepsilon.
	\end{equation*}
with $\mc{Z}_i$ as in \eqref{eq:Z_i}.	\hfill$\square$
\end{definition}
\subsection{Potential game structure}
In this subsection, we prove that  the game  \ref{eq:MPC_second} is an \textit{exact potential game}~\cite{monderer:shapley:1996:potential_games}.
Potential games are characterized by the existence of a potential function that describes the variation of  the cost when an agent changes strategy.
\begin{definition}[Exact potential function]\label{def:potential_game}
	A continuous function $P:\R^n \to \R$ is an \textit{exact potential function} for the game \eqref{eq:MPC_second} if, for all $i \in \mc{I}(k)$, and  $z_i$, $y_i \in \mc{Z}_i(\bs{z}_{-i}|k)$, it satisfies
	\begin{equation*}
	\label{eq:def_potential}
	 J_i(z_i,\bs{z}_{-i}|k) - J_i(y_i,\bs{z}_{-i}|k) =  P(z_i,\bs{z}_{-i}) - P(y_i,\bs{z}_{-i}) \:. \quad\square
	\end{equation*}
%
\end{definition}


To find the potential function $P$, we first reorganize the local cost function $J_i(z_i,\bs z_{-i}|k)$ as:
\begin{equation}
J_i(z_i,\bs z_{-i}|k) = \zeta_i(z_i) +\textstyle{ \sum_{j\in\mc I(k)\setminus \{i\}}} \lambda_{i,j} (z_i,z_j)\:,
\end{equation} 
where $\zeta_i$ depends on the local variables only, and $\lambda_{i,j}$ incorporates the cross terms depending on the other players' strategy $z_j$. From~\eqref{eq:xi_game}, we derive that   
\begin{equation}
\label{eq:lambda_i_j}
\lambda_{i,j} (z_i,z_j)\coloneqq\sum_{t\in\mc T(k)} \chi(t)\gamma \, \vartheta_j(t)\vartheta_i(t).
\end{equation}
Thus, $\lambda_{i,j} (z_i,z_j)=\lambda_{j,i} (z_j,z_i)$ meaning that the agents influence each other in a symmetric way. 
In the next statement, we introduce the exact potential function for the  game in \ref{eq:MPC_second}.
\smallskip
\begin{theorem}\label{prop:potential}
For each $k\in\N$, the game \ref{eq:MPC_second} is an exact potential game with 
\begin{equation*}
\label{eq:pot_function}
P(\bs z|k ) \coloneqq \textstyle{\sum_{i\in\mc I (k)}}\bigg( \zeta_i(z_i)+\sum_{j\in\mc I(k),\, j<i} \lambda_{i,j}(z_i,z_j) \bigg)\,,
\end{equation*}
as an exact potential function, where $\lambda_{i,j}$ is as in \eqref{eq:lambda_i_j}.\hfill $\square$ 
\end{theorem}

\begin{IEEEproof}
The proof is akin to the one in~\cite{fabiani2019nash}.
\end{IEEEproof}

The pivotal result that highlights the importance of the above theorem is that an $\varepsilon$-approximated minimum of the potential function is also an \gls{MINE} of the game \ref{eq:MPC_second}, see \cite[Th.~2]{sagratella2017algorithms}. Thus, it is sufficient to show that the proposed algorithm converges to a minimum of the potential function in order to achieve the sought convergence  result.

\section{\gls{CTM} traffic control scheme}
\label{sec:sol_decision_proc}
We can now focus on the connection between the traffic dynamics and the decision process of the \glspl{PEV}. Then, we describe in details our proposed algorithm that the agents can use to seek an equilibrium of the game.
\subsection{Iterative semi-decentralized algorithm} 
\label{sec:alg}
We propose here a semi-decentralized  iterative algorithm (Algorithm~\ref{alg:Alg_GS}) that the agents in $\mc I(k)$ can adopt to solve the \gls{MI-GPG} \ref{eq:MPC_second}. The notation $z_i(\tau)$ denotes the strategy of agent $i$ at the $\tau$-th iteration of the algorithm. 

After the initialization step, where the players receive the information broadcast by the \gls{HO}, each \gls{PEV} decides to update its strategy independently from the others. If an agent wants to update, it sends a request to the \gls{HO}. If no other player is currently updating, then agent $i$ starts its local update given the aggregate quantities  $\sum_{j\in\mc I(k)\setminus \{i\}} \vartheta_j(t)$, $\sum_{j\in\mc I(k)\setminus \{i\}} u_i(t)$ and $\sum_{j\in\mc I(k)\setminus \{i\}} \delta_i(t)$, used to compute the cost $J_i^{\tup{time}}$ and the coupling constraints \eqref{eq:max_energy_supply}, \eqref{eq:numb_plugs}. On the other hand, if another agent is performing the update, agent $i$ enters a FIFO  queue from which the \gls{HO} extracts sequentially the future agents that are allowed to update. 
At the moment of the update, agent $i$ computes a best-response strategy $z_i^*$ w.r.t. the strategies of the others.
We define the \textit{mixed-integer best-response} mapping for agent $i\in\mc I(k)$, as 
\begin{equation}
\label{eq:BR_mapping}
\mc{BR}_i(\bs z_{-i}) \coloneqq \begin{cases}
&\argmin_{z_i} J_i(z_i, \bs{z}_{-i}|k) \,,\\
&\quad \mathrm{s.t.} \: (z_i,\bs z_{-i}|k)\in \bs{\mc Z}(k)
\end{cases}
\end{equation}
where $\mc{BR}_i$ may be a set, thus $z_i^*\in\mc{BR}_i(\bs z_{-i})$.

 Agent $i$ updates its current strategy only if $z_i^*$  leads to an (at least) $\varepsilon$-improvement in terms of minimization of its cost. 

The iteration is completed after the \gls{PEV} communicates to the \gls{HO} its (possibly) new strategy and the \gls{HO}  uses it to revise all the quantities in the game that  depend on $z_i$.

In the following result, we show that Algorithm~\ref{alg:Alg_GS} converges to an \gls{MINE} of the game \ref{eq:MPC_second}, under the assumption that all the players manage to update their strategies over a sufficiently large number of iterations.
\smallskip
\begin{proposition}
Let $\varepsilon>0$ and $k\in\N$, and assume that for every $j\in\mc I(k)$ and $\tau\in\N$ there exists a $\bar \tau> \tau$ such that $j\in\{i(\tau),\dots,i(\bar \tau)\}$. Then, Algorithm~\ref{alg:Alg_GS} computes an \gls{MINE} of the game \ref{eq:MPC_second} in \eqref{eq:game_compact}. \hfill \QEDopen
\end{proposition} 
\begin{IEEEproof}
From Theorem~\ref{prop:potential}, \ref{eq:MPC_second} is an \gls{MI-GPG} with an exact potential function for all $k\in\N$. Therefore, the result in \cite[Th.~4]{sagratella2017algorithms} applies to show that the sequential best-response based algorithm proposed in Algorithm~\ref{alg:Alg_GS} converges to an \gls{MINE} of the game.  
\end{IEEEproof}

\begin{remark}[Privacy and scalability]
In Algorithm~\ref{alg:Alg_GS}, the \gls{HO} shares with each \gls{PEV} only aggregate information on the choices of the others. This feature allows to preserve the privacy of the agents in the game, since an agent cannot retrieve the local decision strategy of another \gls{PEV} based on the data received from the \gls{HO}. Moreover, using aggregate information is also important to preserve the scalability of Algorithm~\ref{alg:Alg_GS}. In fact, the amount of data shared between each \gls{PEV} and the \gls{HO} does not grow with $n(k)$. This is  crucial  to obtain an implementable solution, due to the (possibly) large number of vehicles involved.
\end{remark}
\setlength{\algomargin}{.5em}
\begin{algorithm}[!t]
	\caption{Sequential best-response}\label{alg:Alg_GS}
	\DontPrintSemicolon
	\SetArgSty{}
	\SetKwIF{If}{ElseIf}{Else}{if}{}{else if}{else}{end if}
	\SetKwFor{ForAll}{for all}{do}{end forall}
	\SetKwRepeat{Do}{do}{end}
	\textbf{Initialization: }For $k\in\N$, \gls{HO} sends to every $i\in\mc I(k)$ the coefficients  $\bar h,u^{\tup{max}},\bar \delta, \gamma\in\R$ and $\xi(t)$, $\vartheta^{\tup{old}}(t)$,$u^{\tup{old}}(t)$,$\delta^{\tup{old}}(t)$, $\hat p(t)$, $\forall t\in \mc T(k)$.\\
\smallskip	
	\textbf{Update: }Choose $\bs{z}(0) \in \bs{\mc{Z}}(k)$, set $\tau \coloneqq 0$\;
	\While{$\bs{z}(\tau)$  is not an \gls{MINE}}{
		\gls{HO} \Do{}{
			Extracts from the waiting queue $i \coloneqq i(\tau) \in \mathcal{I}(k)$.\\
			Sends $\sum_{j\in\mc I(k)\setminus \{i\}} \vartheta_j(t)$, $\sum_{j\in\mc I(k)\setminus \{i\}} u_i(t)$ and $\sum_{j\in\mc I(k)\setminus \{i\}} \delta_i(t)$ to $i$.	
		}
		Player $i$ \Do{}{
			Computes $z^{\ast}_{i}(\tau) \in \mc{BR}_i(\bs z_{-i}(\tau))$ as in \eqref{eq:BR_mapping}\\
			 \uIf{$J_{i}(z_{i}(\tau),\bs z_{-i}(\tau))) - J_{i}(z_{i}^*(\tau),\bs z_{-i}(\tau))) \geq \varepsilon$}{ \smallskip $z_{i}(\tau+1) \coloneqq z^{\ast}_i(\tau)$ 		 
			 \smallskip}		
			 \Else{ $z_{i}(\tau+1) \coloneqq z_i(\tau)$
			 \smallskip}
Sends $z_{i}(\tau+1)$ to \gls{HO}\\ 
		}

		Set $z_j(\tau+1) \coloneqq z_j(\tau) \; \forall j\neq i$, $\tau \coloneqq \tau+1$
}
\end{algorithm}
\subsection{Complete \gls{CTM} control loop}
The  \gls{HO}, introduced in Section~\ref{sec:decision_making_proc},  plays a crucial role in collecting and broadcasting information from and to the vehicles on the highway stretch.
 We propose the following decision process which takes place at the beginning of every time interval $k\in\N$ via the following four  steps.
\subsubsection*{S.1) \gls{HO} collects information}
The \gls{HO} collects information, from the sensors on the highway  (placed at the interfaces between cells), on the cells' density, i.e., $\rho_\ell(k)$ for all $\ell\in\mc N$. The \gls{HO}  computes the following set of variables:  $\xi(t)$ via \eqref{eq:xi_cell}, $\vartheta^{\tup{old}}(t)$, $\delta^{\tup{old}}(t)$, $u^{\tup{old}}(t)$, $\Delta_\ell(k)$ , $\hat \Delta_\ell(t)$ via \eqref{eq:delta_hat}, $p(k)$ via \eqref{eq:price} and $\hat p(t)$ via \eqref{eq:price_hat}, by exploiting the \gls{CTM} and the strategies of the \glspl{PEV} that performed the process during the previous time intervals.

\subsubsection*{S.2) \gls{HO} broadcasts information}
Those \glspl{PEV} that have the possibility to stop during the time interval $k$, i.e., the ones leaving cell $1$, connect with the \gls{HO}, forming the  set $\mc I(k)$ of players involved in the game. The \gls{HO} broadcasts to all of them the quantities they need to initialize the game \ref{eq:MPC_second}, i.e., the initialization phase in Algorithm~\ref{alg:Alg_GS}. Moreover, the \gls{HO} applies the price $p(k)$ in \eqref{eq:price} to the energy purchased by the \glspl{PEV} currently charging at the \gls{CS}.

\subsubsection*{S.3) Iterative solution of the decision process}
After the initialization, the agents update their strategy as shown in Algorithm~\ref{alg:Alg_GS}, and described in Section~\ref{sec:alg}.
 The \glspl{PEV} keep updating  until they converge to an \gls{MINE} of the game \ref{eq:MPC_second}, hence a feasible set of  strategies $\bs z\in\bs{\mc Z}(k)$, which is convenient to each of the \glspl{PEV}. 
 We stress that the iterations $\tau\in\N$ to solve the game are unrelated to the intervals of the \gls{CTM} or the intervals in $\mc T(k)$, and in fact they are all completed within the interval $k$. 

 \subsubsection*{S.4) Strategy implementation}
The agents in $\mc I(k)$ implement their final strategies (i.e., stop at the \gls{CS} or continue driving) and the process will start again from \textit{(S.1)} at the beginning of the interval $k+1$.
\medskip

The presence of the human in the loop imposes a bi-level implementation of step (S.4). We envision that every \gls{PEV} performs the computations in (S.3) via a dedicated software, then the final strategy is translated into a simple message that is prompted to the human user advising whether it is convenient or not to stop at the \gls{CS}. In the end, the driver implements the suggested behavior.

Finally, we want to elaborate on how to compute, starting from $\bs z$,  the in and out flow of the \gls{CS}, i.e., \gls{r2s} and \gls{s2r} respectively. From the constrains in Sec.~\ref{sec:local_coup_constraints}, agent $i\in\mc I(k)$ does not stop at the \gls{CS} if and only if $\vartheta_i(k)=1$, thus the flow entering the \gls{CS} is defined by
\smallskip
\begin{equation}
\label{eq:def_r2s}
\text{\gls{r2s}}(k) \coloneqq  \dfrac{1}{T}\left(n(k)-  \sum_{i\in\mc I(k)} \vartheta_i(k) \right) \:. 
\end{equation}

The flow exiting the \gls{CS} is 
\smallskip
\begin{equation}
\label{eq:def_r2s}
\text{\gls{s2r}}(k) \coloneqq  \dfrac{1}{lT}\sum_{\bar k<k} \sum_{j\in\mc I (\bar k)}  \vartheta_j(k-W)\vartheta_j(k+W) \:, 
\end{equation}
where the double summation selects only those agents exiting the \gls{CS} during the time interval $k$. In fact, for those \glspl{PEV} that do not stop, we have $\sum_{t\in\N}\vartheta_j(t)=W+1$, thus they never contribute to \eqref{eq:def_r2s}. Conversely, if \gls{PEV} $i$ stops,  $\vartheta_j(k-W)\vartheta_j(k+W)=1$ if $i$  exists the \gls{CS} during $k$.
%
%
The \glspl{PEV} that contribute to $\text{\gls{s2r}}(k)$ decide to exit during an interval $m$ long $lT$ time instants that encloses $k$, so  the contribution to the flow is $1/l$. 
The definitions above represent the actual connections  between the \gls{CTM} and the decision process, thus we have arrived at the goal stipulated at the beginning of the paper.
\section{Conclusion}
On a highway stretch with one charging station, the adoption of a dynamic energy price, discounted proportionally to the traffic level, can contribute to alleviating the traffic congestion. It incentivizes the owners of plug-in electric vehicles  to stop for charging during, or close to, rush hours. Under the assumption of a rational self-interested behavior, a multi-agent game arises between the plug-in electric vehicles that have to choose whether or not it is convenient to stop at the charging station. 
The decision process can be formalized as a mixed-integer generalized potential game, and solved via a semi-decentralized iterative scheme, where the highway operator acts as an aggregator. 
 This control scheme converges to a mixed-integer Nash equilibrium of the game, i.e., an approximated optimal charging strategy for the electric vehicles that alleviates the traffic congestion. 
  The effectiveness of our methodology is shown in \cite{cenedese:2020:highway_control_pII} via numerical simulations with real-world data.

This work is the first that proposes a tight integration of traffic dynamics and charging incentives. For this reason, it may be the cornerstone of several further developments. Some assumptions may be relaxed, 
and the same idea can be applied to other  traffic models, 
considering also agents that are not perfectly rational drivers, e.g. using   relative best response dynamics~\cite{cenedese:2019:RBR}. Finally, 
 the case of 
 several charging stations and ramps 
 is left as future work.

\bibliographystyle{IEEEtran}
\bibliography{19_CDC_PEV_MC}

\appendix
\subsection{Translate logical constraints into mixed-integer affine constraints}
First, we show how to translate the basic types of logical implications in sets of inequalities by means of auxiliary variables. Given a linear function $f:\R \rightarrow \R$, let us define $M \coloneqq \textrm{max}_{x \in \mc{X}} f(x)$, $m \coloneqq \textrm{min}_{x \in \mc{X}} f(x)$ where $\mc{X}$ is a  compact set. Then, for $c \in \R$ and $\varphi \in \mathbb{B}$, the triplet $f,\,\varphi,\,c$ satisfy the implication $[\varphi = 1] \iff [f(x) \geq c]$ if and only if it is a solution of the set of mixed-integer inequalities $\mc{S}_{\geq}$, defined as 
\begin{equation*}
\mc{S}_{\geq}(\varphi, f(x), c) \coloneqq \left\{
\begin{aligned}
&(c - m)\varphi \leq f(x) - m\\
&(M - c + \epsilon) \varphi \geq f(x) - c + \epsilon .
\end{aligned}
\right.
\end{equation*}
The value of $\epsilon > 0$ represents a small tolerance on the  constraint violation. Similarly, the condition $[\varphi = 1] \iff [f(x) \leq c]$ is translated into:
\begin{equation*}
\mc{S}_{\leq}(\varphi, f(x), c) \coloneqq \left\{
\begin{aligned}
&(M - c) \varphi \leq M - f(x)\\
&(c + \epsilon - m) \varphi \geq \epsilon + c - f(x).
\end{aligned}
\right.
\end{equation*}

The logical AND between two binary variables $\sigma,\tau\in\B$, i.e,  $[\varphi = 1] \iff [\sigma = 1] \wedge [\tau = 1]$ is equivalent to:
\begin{equation*}
\mc{S}_{\wedge}(\varphi, \sigma, \tau) \coloneqq \left\{
\begin{aligned}
- &\sigma + \varphi \leq 0\\
- &\tau + \varphi \leq 0\\
&\sigma + \tau - \varphi \leq 1,
\end{aligned}
\right.
\end{equation*}

Finally, the product of binary and continuous variables can be transformed into mixed-integer linear inequalities as follows:
\begin{equation*}
\mc{S}_{\Rightarrow}(g, f(x), \varphi) \coloneqq \left\{
\begin{aligned}
& m \varphi \leq g \leq M \varphi\\
& - M(1-\varphi) \leq	g - f(x) \leq - m(1 - \varphi)\\
\end{aligned}\right.
\end{equation*}
The latter is equivalent to: $[\varphi = 0] \implies [g = 0]$, while $[\varphi = 1] \implies [g = f(x)]$.



\subsubsection{Translate the logical implications in~\eqref{eq:MPC_first} }
The expression in \eqref{eq:logical_delta_u} can be directly transformed into the set of linear inequalities
\smallskip
\begin{equation}
\label{eq:log2lin_2}
 \mc{S}_{\geq}(\delta_i(t), u_i(t), \underline{u})\:.
\end{equation}
The two conditions in \eqref{eq:varth_consecutive} can be recast using the two auxiliary variables $\varphi_i^{\tup{LH}},\varphi_i^{\tup{HL}}\in\B$ denoting the rising and falling edge of $\vartheta_i$ respectively, so, for every $t\in\mc T(k)$, they have to satisfy
\begin{subequations}
\label{eq:log2lin_3}
\begin{equation}
{\mc S}_{\wedge}(\varphi_i^{\tup{LH}}(t),(1-\vartheta_i(t-1)),\vartheta_i(t))\:,
\end{equation}
\begin{equation}
{\mc S}_{\wedge}(\varphi_i^{\tup{HL}}(t),\vartheta_i(t-1),(1-\vartheta_i(t)))\:.
\end{equation}
\end{subequations}
Using the variable introduced above for all $t\in\mc T(k)$, the condition in~\eqref{eq:consecutive_intervals} becomes  
\begin{align}
\label{eq:log2lin_pre4a}
&\varphi_i^{\tup{HL}}(t+1) g_i(t)  =0 \:,\\ \nonumber
& g_i(t) \coloneqq \sum_{p\in \mc T(k)}  \vartheta_i(p)-\min\{t-k+W+1, 2W+1\}\:.
\end{align}
The nonlinearity in \eqref{eq:log2lin_pre4a} can be converted using the auxiliary variable $q_i(t)$ for every $t\in\mc T(k)$ as 
\begin{equation}
\label{eq:log2lin_5}
\mc S_{\Rightarrow} (q_i(t),g_i(t),\varphi_i^{\tup{HL}}(t+1))\:.
\end{equation}
Finally, \eqref{eq:log2lin_pre4a} becomes
\begin{equation}
\label{eq:log2lin_6}
q_i(t)=0\:.
\end{equation}
 
The constraint  in \eqref{eq:cnd_no_stop} is equivalent to the linear inequality
\begin{equation}
\label{eq:log2lin_7}
\sum_{p=k}^{t+W+1}\varphi_i^{\tup{HL}}(p) + \delta_i(t) \leq 1\:,\:\forall t\in\mc T(k) \,.
\end{equation}


The logical implication \eqref{eq:exit_CS} requires the introduction of an auxiliary binary variable $\psi_i(t)\in\B$ defined for every $t\in\mc T(k)$ as $[\psi_i(t) = 1] \iff \left[ \delta_i(t-1)=1 \right] \wedge \left[ \delta_i(t)=0  \right]$. This implication can be transformed to the set of inequalities  
\smallskip
\begin{equation}
\label{eq:log2lin_8}
\mc{S}_{\wedge}(\psi_i(t), \delta_i(t-1), (1-\delta_i(t)))\:.
\end{equation}
Then, \eqref{eq:exit_CS} translates into $ [\psi_i(t) = 1] \implies [\vartheta_i(r) =1]$ that can be rephrased, for all $t\in\mc T(k)$ and $r \in\{\max(k,t-W),\dots,t+W\}$, as 
\smallskip
\begin{equation}
\label{eq:log2lin_9}
\vartheta_i(r)-\psi_i(t)\geq 0\:.
\end{equation}
The constraint in \eqref{eq:cnd_charge4h} requires an additional step. First, we introduce the binary auxiliary variable $[\sigma_i(t) = 1] \iff \left[ \delta_i(t-1)=0 \right] \wedge \left[ \delta_i(t)=1  \right]$, for every $t\in\mc T(k)$. By exploiting $\sigma_i$, \eqref{eq:cnd_charge4h} can be equivalently written as
\smallskip
\begin{equation}
\label{eq:cnd_charge4h_recast}
\sigma_i(t) \left( \sum_{h=1}^{\overline{h}} \delta_i(t+h)-\overline{h}  \right) = 0\:.
\end{equation}
Next, to eliminate the nonlinearity in  \eqref{eq:cnd_charge4h_recast}, we define $\overline{h}$ auxiliary binary variables $\mu_i^{(1)}(t),\dots ,\mu_i^{(\overline{h})}(t)\in\B$ for every $t$ as
\smallskip
\begin{equation}
\label{eq:log2lin_10}
\mc{S}_{\wedge}(\mu_i^{(h)}(t), \sigma_i(t), \delta_i(t+h))\,,\: \forall h\in\until{\overline{h}}.
\end{equation}
Thus, \eqref{eq:cnd_charge4h_recast} reduces to the linear equation
\smallskip
\begin{equation}
\label{eq:log2lin_11}
\sum_{h=1}^{\overline{h}} \mu_i^{(h)}(t)-\sigma_i(t)\overline{h}   = 0\:.
\end{equation}
We transform \eqref{eq:min_time_CS} into the following set of linear constraints by introducing the auxiliary variable $\nu_i\in\B$, and hence for every $i\in\mc I(k)$ it must be true that
\smallskip
\begin{equation}
\label{eq:log2lin_11_bis}
\begin{split}
0<  \nu_i  + \tfrac{1}{W+1} \sum_{p=0}^W  \vartheta_i(k+p)\leq 1 \\
\mc S_{\geq} (\nu_i, \textstyle{ \sum_{t\in\mc T(k)} } \vartheta_i(t), W+2 )
\end{split}
\end{equation}

The last logical implication to be transformed into a linear inequality is \eqref{eq:min_charge}. The variable $\omega_i(t)\in\B$ is defined as $[x_i(t)<x_i^{\tup{ref}}] \iff [\omega(t)=1]$, so it satisfies the pattern of inequalities
\smallskip
\begin{equation}
\label{eq:log2lin_12}
\mc{S}_{\leq}(\omega_i(t), x_i(t), x_i^{\tup{ref}})\:.
\end{equation}
Finally, \eqref{eq:min_charge} is equivalent to the linear inequality
\smallskip
\begin{equation}
\label{eq:log2lin_13}
\omega_i(t)+\vartheta_i((\max(k,t-W))) \leq 1\:.
\end{equation}

%
  \begin{IEEEbiography}[{\includegraphics[trim=65 0 65 0,width=1in,height=1.25in,clip,keepaspectratio]{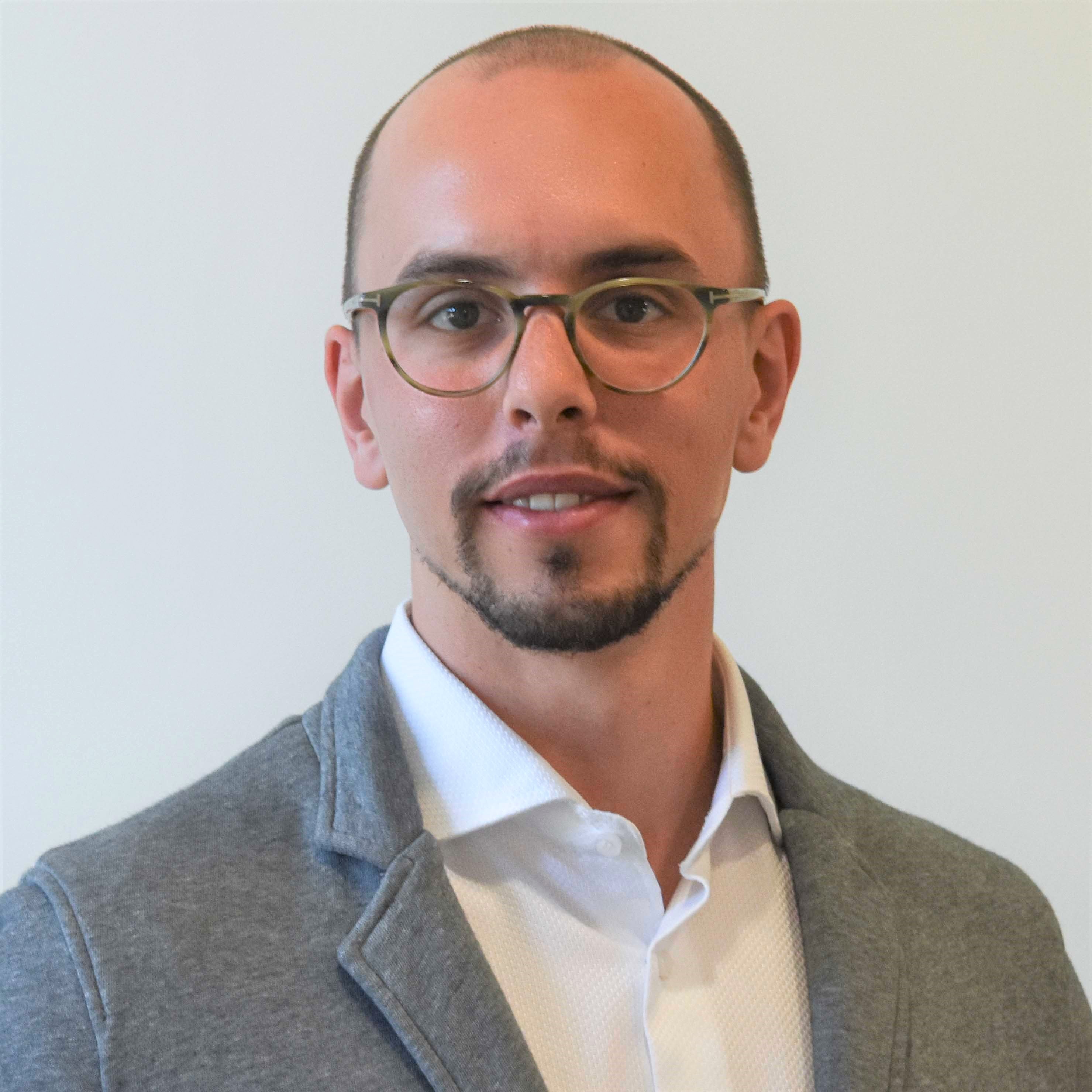}}]{Carlo Cenedese}
received the M.Sc. degree in Automation Engineering from the University of Padova, Padova, Italy, in 2016. 
From August to December 2016, he worked for VI-grade srl in collaboration with the  Automation Engineering group of the University of  Padova as a research fellow. In 2021, he completed his Ph.D. degree with the Discrete Technology and Production Automation (DTPA) Group in the Engineering and Technology Institute (ENTEG) at the University of Groningen, the Netherlands. During December 2019 he visited the Department of Mathematical Sciences, Politecnico di Torino, Torino, Italy. 
Since 2021 he is a Postdoc in the Department of Information Technology and Electrical Engineering at  the ETH Z\"urich.  
His research interests include game theory, traffic control, complex networks and multi-agent network systems associated to decision-making processes.
\end{IEEEbiography}

\begin{IEEEbiography}[{\includegraphics[trim=0 0 0 0,width=1in,height=1.25in,clip,keepaspectratio]{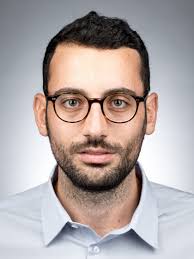}}]{Michele Cucuzzella}
received the M.Sc. degree (Hons.) in Electrical Engineering and the Ph.D. degree in Systems and Control from the University of Pavia, Pavia, Italy, in 2014 and 2018, respectively. Since 2021 he is Assistant Professor of automatic control at the University of Pavia. From 2017 until 2020, he was a Postdoc at the University of Groningen, the Netherlands. From April to June 2016, and from February to March 2017 he was with the Bernoulli Institute of the University of Groningen. His research activities are mainly in the area of nonlinear control with application to the energy domain and smart systems. He has co-authored the book Advanced and Optimization Based Sliding Mode Control: Theory and Applications, SIAM, 2019. He serves as Associate Editor for the European Control Conference since 2018 and received the Certificate of Outstanding Service as Reviewer of the IEEE Control Systems Letters 2019. He also received the 2020 IEEE Transactions on Control Systems Technology Outstanding Paper Award, the IEEE Italy Section Award for the best Ph.D. thesis on new technological challenges in energy and industry, the SIDRA Award for the best Ph.D. thesis in the field of systems and control engineering and he was one of the finalists for the EECI Award for the best Ph.D. thesis in Europe in the field of control for complex and heterogeneous systems.
\end{IEEEbiography}

 \begin{IEEEbiography}[{\includegraphics[width=1in,height=1.25in,clip,keepaspectratio]{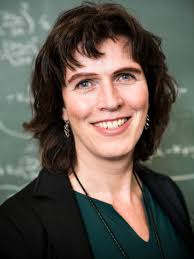}}]{Jacquelien M.A. Scherpen} received the M.Sc.
and Ph.D. degrees in applied mathematics from the
University of Twente, The Netherlands, in 1990 and
1994, respectively. She was with Delft University of
Technology, The Netherlands, from 1994 to 2006.
Since September 2006, she has been a Professor with
the University of Groningen, The Netherlands, at
the Engineering and Technology institute Groningen
(ENTEG) of the Faculty of Science and Engineering.
From 2013 til 2019 she was scientific director of
ENTEG. She is currently Director of the Groningen
Engineering Center. She has held various visiting research positions at
international universities. Her current research interests include model reduction
methods for networks, nonlinear model reduction methods, nonlinear
control methods, modeling and control of physical systems with applications
to electrical circuits, electromechanical systems, mechanical systems, and
grid application, and distributed optimal control applications to smart grids.
Jacquelien Scherpen has been an Associate Editor of the IEEE Transactions on
Automatic Control, the International Journal of Robust and Nonlinear Control
(IJRNC), and the IMA Journal of Mathematical Control and Information. She
is on the Editorial Board of the IJRNC. She was awarded the best paper prize
for Automatica 2017-2020. In 2019 she received a royal distinction and is
appointed Knight in the Order of the Netherlands Lion. Since 2020 she is
Captain of Science of the Dutch topsector High Tech Systems and Materials.
She is council member of IFAC, member of the BoG of the IEEE Control
Systems Society, and president of the European Control Association (EUCA).
\end{IEEEbiography} 
 
\vspace{-2cm}
\begin{IEEEbiography}[{\includegraphics[trim=40 0 0 0,width=1in,height=1.25in,clip,keepaspectratio]{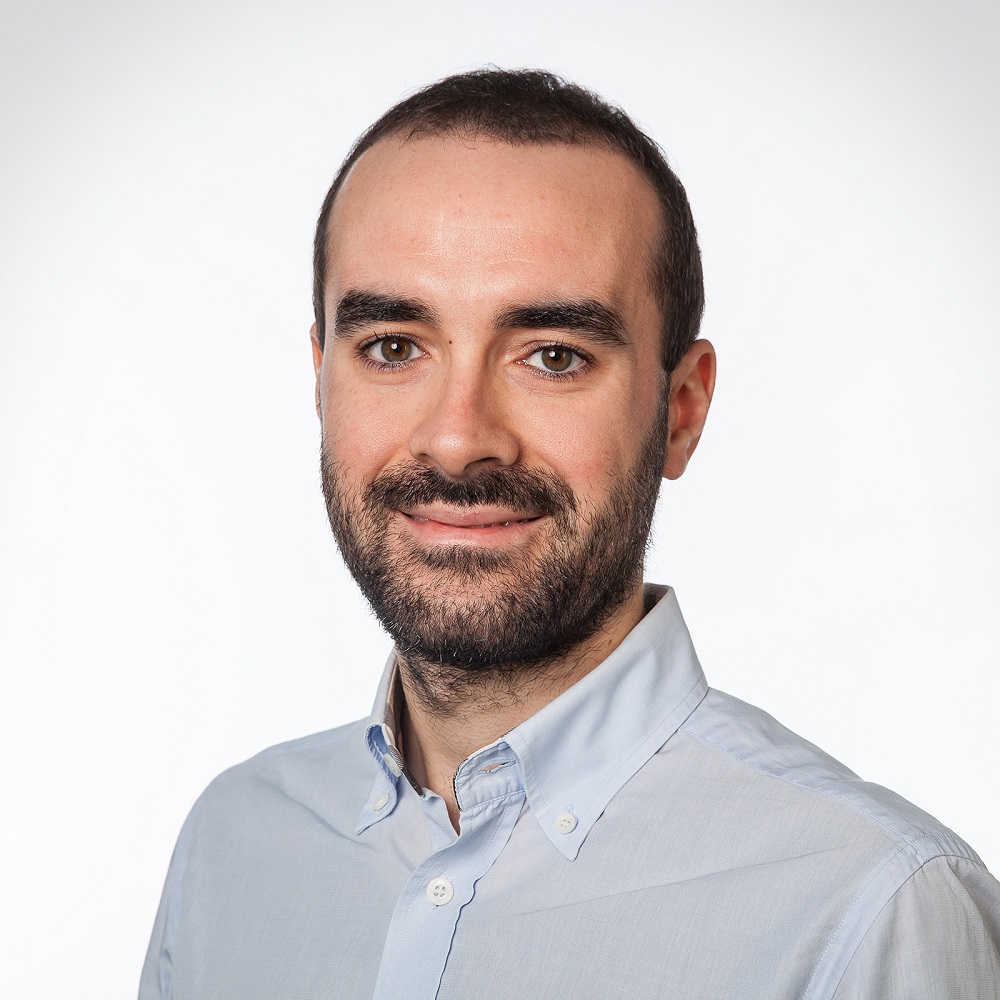}}]{Sergio Grammatico}(M’16 - SM’19) is an associate professor at the Delft Center for Systems and Control, TU Delft, The Netherlands. He received B.Sc., M.Sc. and Ph.D. degrees in Automatic Control Engineering from the University of Pisa, Italy, in 2008, 2009, and 2013 respectively, and an M.Sc. degree in Engineering Science from the Sant’Anna School of Advanced Studies, Pisa, Italy, in 2011. Prior to joining TU Delft, he was a post-doctoral researcher in the Automatic Control Laboratory, ETH Zurich, Switzerland, and an assistant professor in the Department of Electrical Engineering, TU Eindhoven, The Netherlands. He was awarded a 2005 F. Severi B.Sc. scholarship by the Italian High-Mathematics National Institute, and a 2008 M.Sc. fellowship by the Sant’Anna School of Advanced Studies. He was awarded 2013 and 2014 TAC Outstanding Reviewer and he was a recipient of the Best Paper Award at the 2016 ISDG International Conference on Network Games, Control and Optimization. His research interests include distributed optimization and monotone game theory for complex systems of systems. He is currently an Associate Editor for the IEEE Transactions on Automatic Control 34 and Automatica.
\end{IEEEbiography} 

\vspace{-10cm}
\begin{IEEEbiography}[{\includegraphics[trim=80 0 60 0,width=1in,height=1.25in,clip,keepaspectratio]{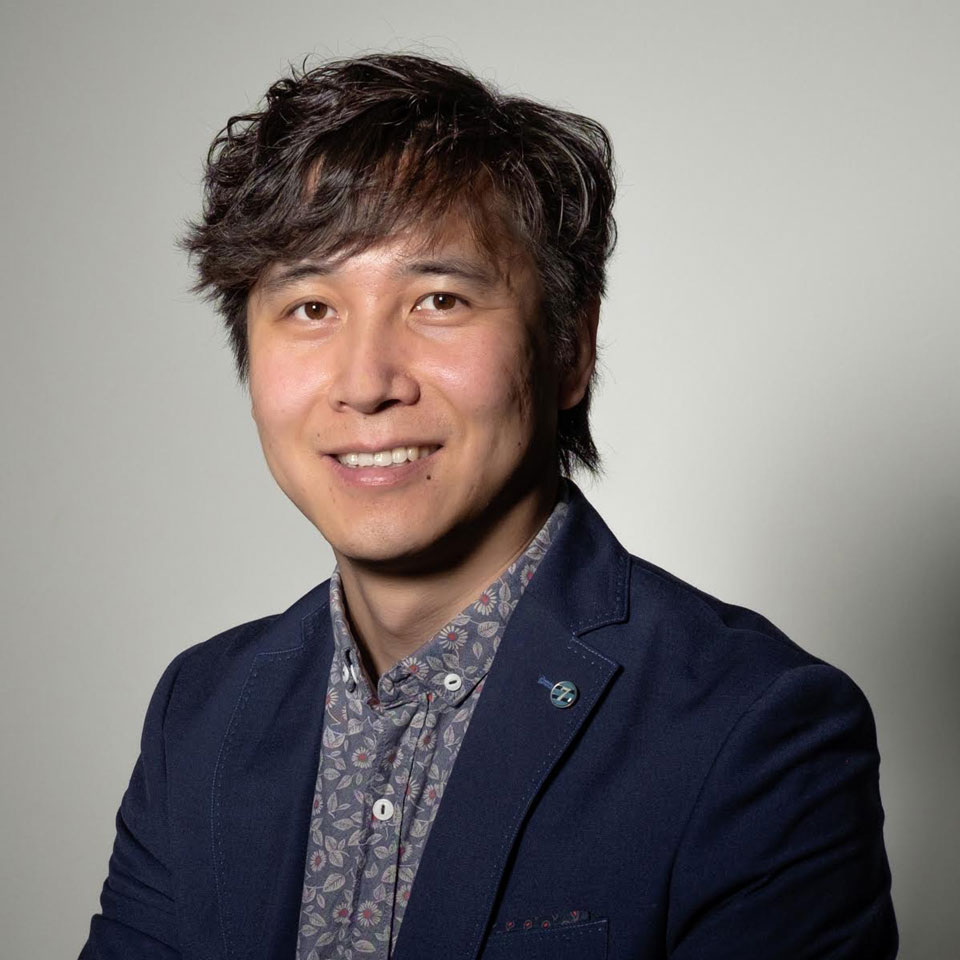}}]{Ming Cao}
has since 2016 been a professor of systems and control with the Engineering and Technology Institute (ENTEG) at the University of Groningen, the Netherlands, where he started as a tenure-track Assistant Professor in 2008. He received the Bachelor degree in 1999 and the Master degree in 2002 from Tsinghua University, Beijing, China, and the Ph.D. degree in 2007 from Yale University, New Haven, CT, USA, all in Electrical Engineering. From September 2007 to August 2008, he was a Postdoctoral Research Associate with the Department of Mechanical and Aerospace Engineering at Princeton University, Princeton, NJ, USA. He worked as a research intern during the summer of 2006 with the Mathematical Sciences Department at the IBM T. J. Watson Research Center, NY, USA. He is the 2017 and inaugural recipient of the Manfred Thoma medal from the International Federation of Automatic Control (IFAC) and the 2016 recipient of the European Control Award sponsored by the European Control Association (EUCA). He is a Senior Editor for Systems and Control Letters, and an Associate Editor for IEEE Transactions on Automatic Control, IEEE Transactions on Circuits and Systems and IEEE Circuits and Systems Magazine. He is a vice chair of the IFAC Technical Committee on Large-Scale Complex Systems. His research interests include autonomous agents and multi-agent systems, complex networks and decision-making processes.  
\end{IEEEbiography}

\end{document}